\documentclass{aa}  

\usepackage{CJK}
\usepackage{xcolor}
\usepackage{graphicx}
\usepackage{txfonts}
\usepackage{lipsum}
\usepackage{hyperref}
\usepackage{footnotehyper}
\usepackage{subcaption}

\begin{document}

   \title{SN~2024igg: A Super-Chandrasekhar/03fg-like SN exhibiting C\,{\sc ii}-dominated spectra after explosion}


\author{
Jialian Liu \inst{\ref{thu}\thanks{E-mail: liu-jl22@mails.tsinghua.edu.cn}}
\and Xiaofeng Wang \inst{\ref{thu}\thanks{E-mail: wang\_xf@mail.tsinghua.edu.cn}}
\and Liyang Chen \inst{\ref{thu}}
\and Alexei V. Filippenko \inst{\ref{Berkeley}}
\and Thomas G. Brink \inst{\ref{Berkeley}}
\and WeiKang Zheng \inst{\ref{Berkeley}}
\and Andrea~Pastorello \inst{\ref{INAFa}}
\and Paolo Ochner \inst{\ref{UdSdP},\ref{INAFa}}
\and Irene Albanese \inst{\ref{UdSdP},\ref{CISAS}}
\and Andrea Reguitti \inst{\ref{INAFa},\ref{INAFb}}
\and Giorgio Valerin \inst{\ref{INAFa}}
\and Yongzhi~Cai \begin{CJK}{UTF8}{gbsn}(蔡永志)\end{CJK} \inst{\ref{YOCA},\ref{ICOS},\ref{KLSEC},\ref{INAFa}}
\and Jujia Zhang \inst{\ref{YOCA},\ref{ICOS},\ref{KLSEC}}
\and Liping Li \inst{\ref{YOCA},\ref{ICOS},\ref{KLSEC}}
\and Zhenyu Wang \inst{\ref{YOCA},\ref{ICOS},\ref{KLSEC},\ref{UCAS}}
\and Liangduan Liu \inst{\ref{CCNU},\ref{NADC},\ref{KLQLP}}
\and Yuhao Zhang \inst{\ref{CCNU},\ref{NADC},\ref{KLQLP}}
}

\institute
{Department of Physics, Tsinghua University, Beijing, 100084, China \label{thu}
\and Department of Astronomy, University of California, Berkeley, CA 94720-3411, USA \label{Berkeley}
\and INAF -- Osservatorio Astronomico di Padova, Vicolo dell'Osservatorio 5, 35122 Padova, Italy \label{INAFa}
\and Universit\`a degli Studi di Padova, Dipartimento di Fisica e Astronomia, Vicolo dell'Osservatorio 2, 35122 Padova, Italy \label{UdSdP}
\and CISAS, Universit\`a degli Studi di Padova, Via Venezia, 15, 35131 Padova, Italy \label{CISAS}
\and INAF -- Osservatorio Astronomico di Brera, Via Bianchi 46, 23807 Merate (LC), Italy \label{INAFb}
\and Yunnan Observatories, Chinese Academy of Sciences, Kunming 650216, China \label{YOCA}
\and International Centre of Supernovae, Yunnan Key Laboratory, Kunming 650216, China \label{ICOS}
\and Key Laboratory for the Structure and Evolution of Celestial Objects, Chinese Academy of Sciences, Kunming 650216, China \label{KLSEC}
\and School of Astronomy and Space Science, University of Chinese Academy of Sciences, Beijing 100049, China \label{UCAS}
\and Institute of Astrophysics, Central China Normal University, Wuhan 430079, China \label{CCNU}
\and Education Research and Application Center, National Astronomical Data Center, Wuhan 430079, China \label{NADC}
\and Key Laboratory of Quark and Lepton Physics (Central China Normal University), Ministry of Education, Wuhan 430079, China \label{KLQLP}
}

   \date{Received January XX, 2026}

 
  \abstract
   {}
{We present and analyze photometric and spectroscopic observations of the Type Ia supernova (SN~Ia) 2024igg, another ``super-Chandrasekhar'' (or 03fg-like) SN whose strong C\,{\sc II} $\lambda6580$ feature was initially misidentified as H$\alpha$, thereby constraining its progenitor system, explosion parameters, and physical scenario.
    } 
{We conducted a comparative analysis of SN~2024igg and other 03fg-like SNe. The radiative transfer code \textsc{TARDIS} was run to identify the features in the first spectrum at $t\approx\ $$-$13.9 days. A pseudobolometric light curve was constructed using ultraviolet and optical photometry and then modeled with fitting tools. 
   }
   {SN~2024igg shows many characteristics in common with other 03fg-like objects, such as high ultraviolet flux, slowly declining light curves ($\Delta m_{15}(B)=0.90\pm0.08$ mag), low expansion velocities, along with strong and persistent C\,{\sc II} absorption. Meanwhile, this SN exhibits some remarkable properties within this subgroup, including a moderately low optical luminosity ($M_{\rm max}(B)=-18.99\pm0.15$ mag), a short rise time less than 18.5 days, and strong C\,{\sc II} $\lambda6580$. The bolometric analysis yields a $^{56}$Ni mass of $M_{\rm Ni}=0.547\pm0.082$ $M_{\rm \odot}$ and an ejecta mass of $1.54^{+0.22}_{-0.19}$ $M_{\rm \odot}$, marginally exceeding the Chandrasekhar mass. Our \textsc{TARDIS} result indicates that most of the features in the earliest spectrum could be attributed to C\,{\sc II}, which is consistent with a model where a supernova explodes within a carbon-rich circumstellar medium (CSM). The CSM interaction would produce a density peak in the ejecta, offering a natural explanation for the slowly evolving line velocities near $-$8000 km s$^{-1}$. The CSM may stem from the debris of a secondary white dwarf in a white-dwarf merger or the envelope of an asymptotic giant branch star. Combined with the unshifted forbidden lines in the spectrum taken at $t\approx\ +$135 days, we suggest that SN~2024igg comes from a symmetric explosion on a secular timescale after the merger.
   }
   {}


   \maketitle

\nolinenumbers
\section{Introduction}\label{sec:Intro}

It is widely accepted that Type Ia supernovae (SNe Ia; see, e.g., \citealt{1997ARA&A..35..309F} for a review on SN classification) originate from thermonuclear explosions of carbon-oxygen (C/O) white dwarfs \citep[WDs;][]{1997Sci...276.1378N,2000ARA&A..38..191H}. Normal SNe Ia, which adhere to the width-luminosity relation (WLR, also dubbed the ``Lira-Phillips relation"; \citealt{1993ApJ...413L.105P}; \citealt{1999AJ....118.1766P}), serve as vital standardizable candles for measuring the distances to their host galaxies and have played a critical role in revealing the accelerating expansion of the Universe \citep{1998AJ....116.1009R,1999ApJ...517..565P} -- though their progenitor systems and explosion mechanisms remain under debate. In recent years, a growing number of peculiar SNe Ia that deviate from the WLR have been observed \citep{2017hsn..book..317T}, offering new opportunities to probe the final stage of stellar evolution and the physics of thermonuclear explosions.

The super-Chandrasekhar-mass (super-$M_{\rm Ch}$) SNe are a peculiar subclass of SNe~Ia. The moniker for this subclass comes from the super-$M_{\rm Ch}$ WD progenitor inferred from the light curve of the prototype SN~2003fg \citep{2006Natur.443..308H}. 
Specifically, powering the overluminous and broad light curve of SN~2003fg requires a $^{56}$Ni mass of $\sim$1.3 $M_{\rm \odot}$. This value is already close to $M_{\rm Ch}$ (1.4 $M_{\rm \odot}$) and thus implies a super-$M_{\rm Ch}$ explosion, since even the pure detonation of a $M_{\rm Ch}$ WD can produce only 0.92 $M_{\rm \odot}$ of $^{56}$Ni \citep{1993A&A...270..223K}. 
As the sample grows, SNe~Ia of this subclass are also found to be characterized by a broad light curve, blue ultraviolet (UV) through optical colors, a weak or no $i$-band secondary maximum, moderately low ejecta velocities, and strong and persistent C\,{\sc ii} absorption \citep{2017hsn..book..317T}. 

Meanwhile, some SNe in common with these characteristics are found to have a comparatively moderate luminosity, similar to \citep[e.g., SN~2012dn;][]{2014MNRAS.443.1663C} or even fainter \citep[e.g., ASASSN-15hy;][]{2021ApJ...920..107L} than a normal SN~Ia with the same light-curve width. An explosion of a super-$M_{\rm Ch}$ WD seems unnecessary for these moderate SNe, and thus hereafter we follow the moniker of ``03fg-like'' proposed by \citet{2021ApJ...922..205A} for this subclass. An interaction of SN ejecta and a carbon-rich circumstellar medium (CSM), which could originate from debris of a double-WD merger \citep{2007MNRAS.380..933Y} or the envelope of an asymptotic giant branch (AGB) star, has been included to interpret the common features of 03fg-like SNe in many studies (e.g., \citealt{2013MNRAS.432.3117T,2021ApJ...920..107L,2021ApJ...922..205A,2022ApJ...927...78D,2023MNRAS.521.1162D,2024ApJ...960...88S}), since it can increase the total ejecta mass, enrich the carbon abundance, and possibly power the light curve, as well as slow down the ejecta velocity \citep{2016MNRAS.463.2972N}. To verify this scenario, an extremely early-time spectrum would be highly valuable, as the interaction could cause the spectrum to be temporarily dominated by carbon and oxygen lines when the photosphere resides in the C/O-rich shell \citep{2023MNRAS.521.1897M}. But spectroscopic observations within days of explosion are available for only a handful of 03fg-like SNe, within which only SN~2020esm \citep{2022ApJ...927...78D} exhibits a spectrum that can be well fit with just carbon and oxygen lines. 

Recently, another 03fg-like object with an early-time spectrum dominated by C\,{\sc II} was discovered: SN~2024igg. Here we present optical photometric and spectroscopic observations of it. The discovery and our observations are described in Section~\ref{sec:Obs}. In Section~\ref{sec:analysis}, we present the photometric and spectroscopic evolution of this SN and compare with other well-observed SNe~Ia. We also perform analysis on the first spectrum, alongside fits to the pseudobolometric light curve. Evidence of CSM interaction and its origin are discussed in Section~\ref{sec:Discussions}. We summarize our findings in Section~\ref{sec:Conclusions}. 

\section{Observations and data reduction}\label{sec:Obs}

\subsection{Discovery}\label{subsec:discovery}

SN~2024igg was discovered \citep{2024TNSTR1413....1M} on 2024 May 7.30 (UTC dates are used throughout this paper; MJD 60437.30) by the Zwicky Transient Facility (ZTF) at $\alpha=15^{\rm hr}09^{\rm m}30.910^{\rm s}$, $\delta=54^\circ30^{\prime}20.24^{\prime\prime}$ (J2000). The last nondetection reported by \citet{2024TNSTR1413....1M} was on MJD 60437.26, just $\sim 0.04$~d before the first detection. But the forced photometry provided by Lasair\footnote{https://lasair-ztf.lsst.ac.uk} indicates a last nondetection on MJD 60433.41 to a limit of 20.86 mag and a first detection of 19.10 mag on MJD 60436.45, both in the ZTF $r$ band. This work adopts the Lasair results. Note that the ATLAS $o$ band has a nondetection on MJD 60435.44, but the limit is only 18.74 mag. 

\begin{figure}
        \centering
        \includegraphics[width=\hsize]{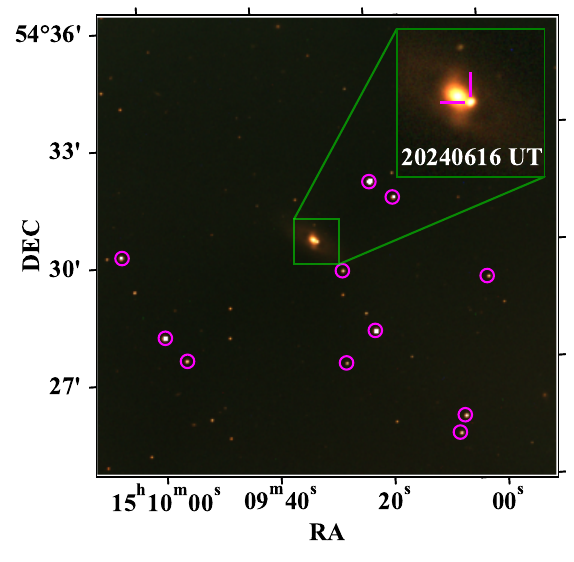}
        \caption{TNT $12' \times 12'$ color-composite ($B/V/r$) image of SN~2024igg and its host galaxy. The reference stars used to calibrate the photometry are marked with circles. The inset shows the zoomed-in region of the SN (indicated with magenta lines) and its host taken on 2024 June 16.}
        \label{fig:position}%
    \end{figure}

In Fig.~\ref{fig:position}, we present a $11.9' \times 11.9'$ color-composite ($B/V/r$) image, taken with the 0.8~m Tsinghua University-NAOC telescope \citep[TNT;][]{2008ApJ...675..626W,2012RAA....12.1585H}, of SN~2024igg and nearby field stars. The zoomed-in view shows that the projected location of 
SN~2024igg is close to the center of the host galaxy NGC 5876, a barred spiral at a redshift of $z=0.01085$ (value from the NASA/IPAC Extragalactic Database, NED\footnote{\url{https://ned.ipac.caltech.edu/}}). Originally identified as a Type II SN by \citet{2024TNSCR1433....1M} based on a spectrum obtained with the Nordic Optical Telescope (NOT), this object was reclassified as an 03fg-like Type Ia SN by \cite{2024TNSAN.132....1S} based on the identifications of C\,{\sc ii} and Si\,{\sc ii} lines in their spectrum taken with the 2~m Liverpool Telescope \citep[LT;][]{2004SPIE.5489..679S}. 

\subsection{Photometry}

We performed photometry of SN~2024igg in the $griBV$ filters with the 0.8~m 
TNT at Xinglong Station of NAOC, the Lijiang 2.4~m telescope \citep[LJT;][]{2015RAA....15..918F} of Yunnan Observatories, and the Schmidt 67/91 cm Telescope (67/91-ST) of the Osservatorio Astronomico di Asiago. 

All of these images were preprocessed to subtract the bias and correct the flat-field. Since the SN is seriously contaminated by the host-galaxy bulge, we performed template subtraction using \textsc{HOTPANTS} \citep{hotpants}. The templates for TNT and 67/91-ST were obtained after the SN had faded. For the LJT images, we performed the subtraction using the TNT templates. Then we applied aperture photometry with the Automated Photometry of Transients pipeline \citep[\textsc{AutoPhOT};][]{2022A&A...667A..62B} for both the local reference stars and the template-subtracted SN. The aperture correction was derived from the local reference stars and applied to the SN photometry. The instrumental magnitudes were calibrated against Gaia synthetic photometry \citep{2023A&A...674A..33G}.

SN~2024igg was also observed by the Ultraviolet/Optical Telescope (UVOT; \citealp{2004ApJ...611.1005G,2005SSRv..120...95R}) onboard  the {\it Neil Gehrels Swift Observatory} \citep{2004ApJ...611.1005G} in three UV ($UVW2$, $UVM2$, $UVW1$) and three optical ($U$, $B$, $V$) filters. We extracted {\it Swift} photometry using \textsc{Swift\_host\_subtraction} \citep{2009AJ....137.4517B,2014Ap&SS.354...89B} with the latest \emph{Swift} calibration database\footnote{\url{https://heasarc.gsfc.nasa.gov/docs/heasarc/caldb/swift/}}. The host flux subtraction was applied to remove the contamination from the host galaxy. In addition, we included the $g$- and $r$-band photometry of ZTF provided by Lasair.

\subsection{Spectroscopy}

Optical spectra of SN~2024igg were collected with different instruments, including the Beijing-Faint Object Spectrograph and Camera (BFOSC) mounted on the Xinglong 2.16~m telescope \citep[XLT;][]{2016PASP..128j5004Z}, the Yunnan Faint Object Spectrograph and Camera (YFOSC) on the LJT, the Asiago Faint Objects Spectrograph and Camera (AFOSC) on the Copernico Telescope on Mount Ekar, the Boller \& Chivens spectrograph (B\&C) on the Galileo Telescope at Osservatorio Astrofisico di Asiago, the Low Resolution Imaging Spectrometer \citep[LRIS;][]{1995PASP..107..375O} on the 10~m Keck-I telescope on Maunakea, the Kast double spectrograph \citep{miller1993lick} on the 3~m Shane telescope at Lick Observatory, and the Low Resolution Spectrograph (LRS) on Telescopio Nazionale Galileo (TNG) on the island of La Palma. 
Standard \textsc{IRAF}\footnote{{IRAF} is distributed by the National Optical Astronomy Observatories, which are operated by the Association of Universities for Research in Astronomy, Inc., under cooperative agreement with the U.S. National Science Foundation (NSF).} routines were used to reduce all the spectra except for the Keck~I/LRIS spectrum, which was reduced using the \textsc{LPipe} pipeline \citep{2019PASP..131h4503P}. Flux calibration of the spectra was performed with spectrophotometric standard stars observed on the same nights. Atmospheric extinction was corrected with the extinction curves of local observatories. 
We derived the telluric correction using the spectrophotometric standard star. LRIS has an atmospheric-disperson corrector to minimize differential slit losses \citep{1982PASP...94..715F}. Most of the spectra from the other facilities were obtained at low airmass to reduce such losses. In addition, we include the two spectra mentioned in Section\,\ref{subsec:discovery}.

All of the photometry and a journal of spectroscopic observations of SN~2024igg are presented in Appendix\,\ref{sec:observations_table}.  


\section{Analysis}\label{sec:analysis}

\subsection{Reddening and distance}\label{subsec:host}
 
The Galactic reddening on the line of sight is $E(B-V)_{\rm Gal}=0.011$ mag \citep{2011ApJ...737..103S}. No obvious interstellar Na\,{\sc i}\,D absorption at the host redshift is seen in the spectra of SN~2024igg, although the projected distance of the SN to its host-galaxy center is small. We therefore adopt a reddening of $E(B-V)=0.011$ mag and a \citet{1999PASP..111...63F} reddening law with $R_V=3.1$ in this work.
The sole redshift-independent distance estimate for NGC 5876 comes from the Tully-Fisher relation, yielding a distance modulus of $34.08\pm0.46$ mag \citep{2007A&A...465...71T} with a large uncertainty. We use the cosmological distance in this work, as the redshift of NGC 5876 is $\gtrsim0.01$. Adopting H$_0 = 73\pm5$ km s$^{-1}$ Mpc$^{-1}$ (the value is taken from \citealt{2022ApJ...934L...7R} and the uncertainty considers the Hubble tension) and correcting for peculiar motions related to the Virgo cluster and Great Attractor \citep{2000ApJ...529..786M}, we estimate the distance to NGC 5876 to be $51.2\pm3.5$ Mpc. This corresponds to a distance modulus of $33.55\pm0.15$ mag.

\subsection{Photometric evolution}\label{subsec:LC_evolution}

\begin{figure*}
        \centering
        \includegraphics[width=0.49\textwidth]{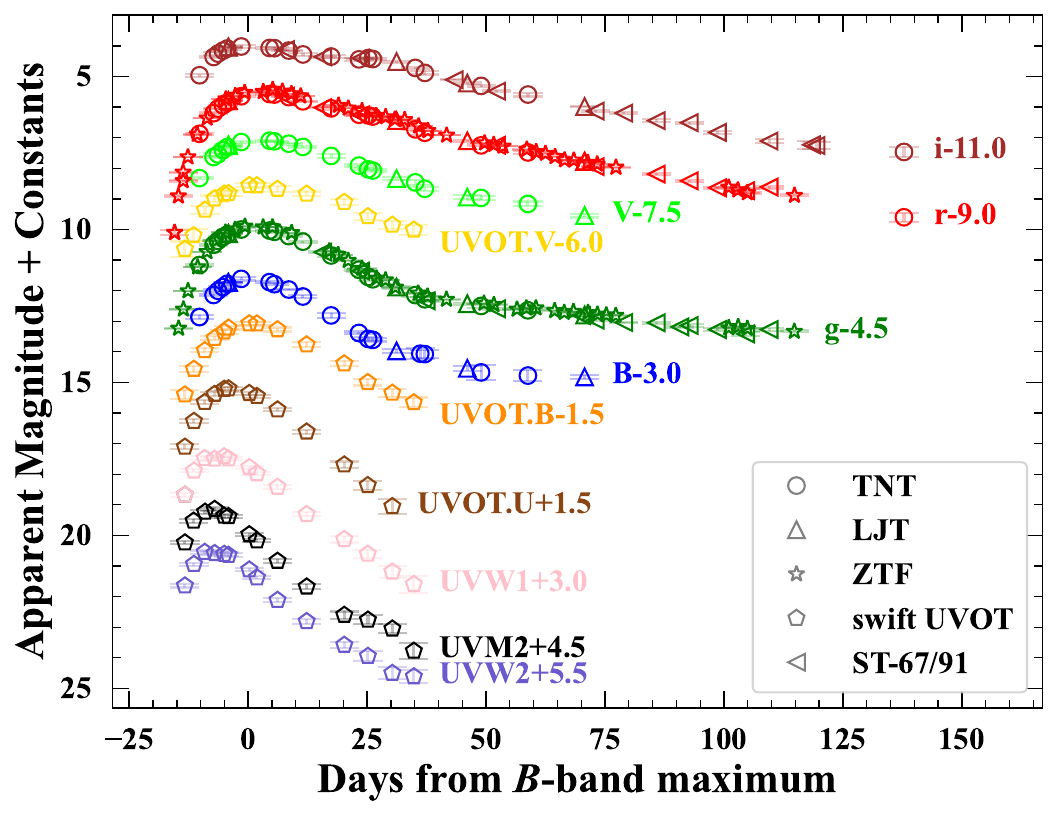}
        \includegraphics[width=0.49\textwidth]{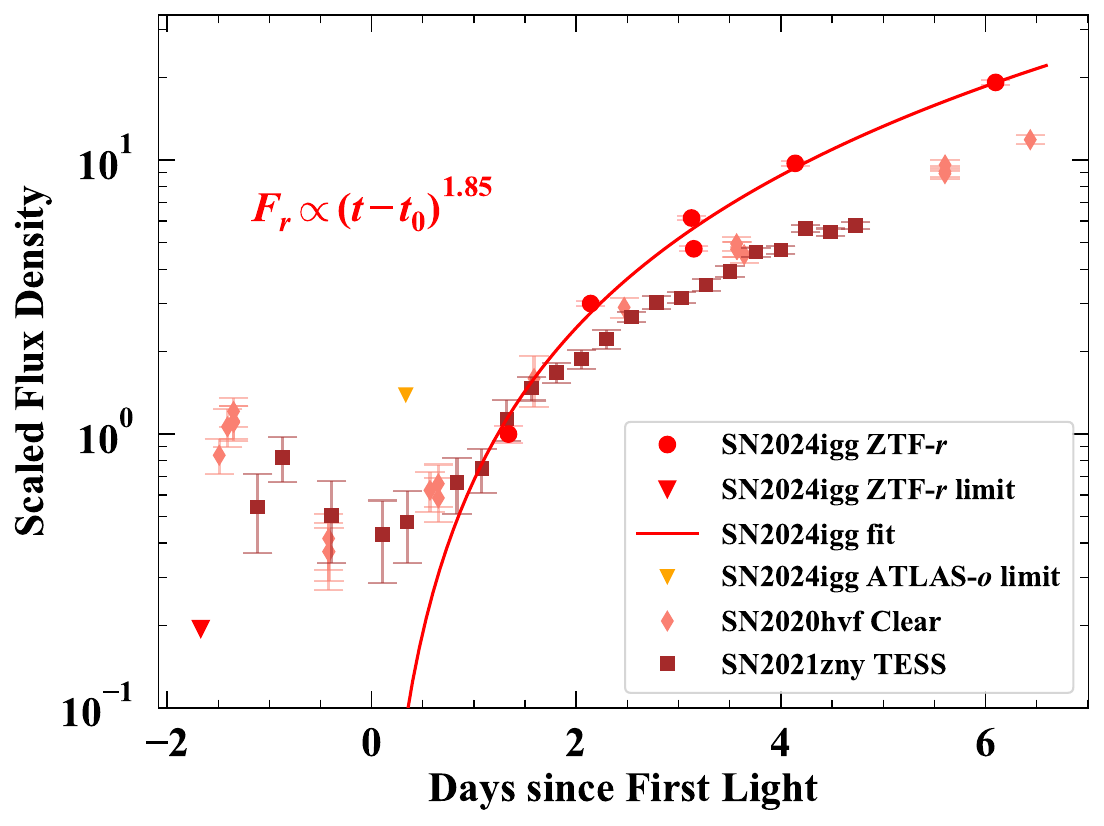}
        \caption{
        {\it Left panel:} UV and optical light curves of SN~2024igg. The phase is with respect to the $B$-band maximum (MJD = 60452.10; see Section\,\ref{subsec:LC_evolution}). Data in the different filters are shown with different colors, and are shifted vertically for better display. The corresponding instruments are indicated in the legend. {\it Right panel:} Power-law fit to the early-time ZTF $r$-band light curve of SN~2024igg. The $r$-band data of SN~2024igg used to fit the power law are shown in red circles. The last nondetections from ZTF $r$ and ATLAS $o$ are plotted as inverted triangles. For comparison, the Clear-band data of SN~2020hvf and the TESS observations of SN~2021zny are shown as salmon diamonds and brown squares, respectively. The flux densities of SN~2024igg are converted from observed magnitudes and are then scaled to the first detection point. For SNe~2020hvf and 2021zny, the phases are shifted to align their explosion time to the last nondetection of SN~2024igg in ZTF $r$, and their flux densities are scaled according to the power-law curve around the epoch of the first detection of SN~2024igg.}
        \label{fig:all_lc}%
    \end{figure*}


Fig.~\ref{fig:all_lc} shows all the light curves of SN~2024igg. A fourth-order polynomial fit was applied to the $B$-band light curve of SN~2024igg around maximum light, with which we found a peak of $B_{\rm max}=14.61\pm0.03$ mag on 2024 May $22.10 \pm 0.42$ (MJD = $60452.10 \pm 0.42$) with $\Delta m_{15}(B)=0.90\pm0.08$ mag. Assuming $R_B=4.1$ and a reddening of $E(B-V)=0.011$ mag, the $B$-band absolute peak magnitude is $M_{\rm max}(B)=-18.99\pm0.15$ mag.

To estimate the time of first light, we tried to fit a power law to the ZTF $r$-band rise phase of the light curve using the data points where the SN flux was $\lesssim 40$\% of its peak value,
\begin{equation}
F_r=A(t-t_0)^\alpha ,
\label{eq:power_law}
\end{equation}
where $F_r$ is the flux in the $r$ band, $A$ is the scale factor, $t_0$ is the time of first light, and $\alpha$ is the power-law index. The best fit is shown in the right-hand panel of Fig.~\ref{fig:all_lc}. We found that a simple power law, with $t_0=60435.09\pm0.16$ (MJD) and $\alpha=1.85\pm0.09$, could roughly describe the rise points. But this $t_0$ corresponds to a rise time of only about 17 days to $B$-band maximum, much shorter than the average rise time of $22.0\pm3.8$ days of 03fg-like SNe \citep{2021ApJ...922..205A}. This might be attributed to a dark phase between the explosion and the first light powered by inner radioactive material. Note that, even when adopting the last nondetection as the explosion epoch, the rise time is still less than 18.5 days, possibly shorter than that of any other 03fg-like SN. An early excess that deviates from a simple power-law rise is expected if there is an interaction between the SN ejecta and CSM, which has been observed in several 03fg-like SNe such as SNe~2020hvf \citep{2021ApJ...923L...8J} and 2021zny \citep{2023MNRAS.521.1162D}. 
No early excess is detected in SN~2024igg, but this might be caused by  the brief ($\sim$1 day) excess phase falling between observational gaps. Specifically, if we shift the phases of SNe~2020hvf and 2021zny to align their explosion times to the last nondetection of SN~2024igg, their early excess would have faded before the first detection of SN~2024igg.

\begin{figure}
        \centering
        \includegraphics[width=\hsize]{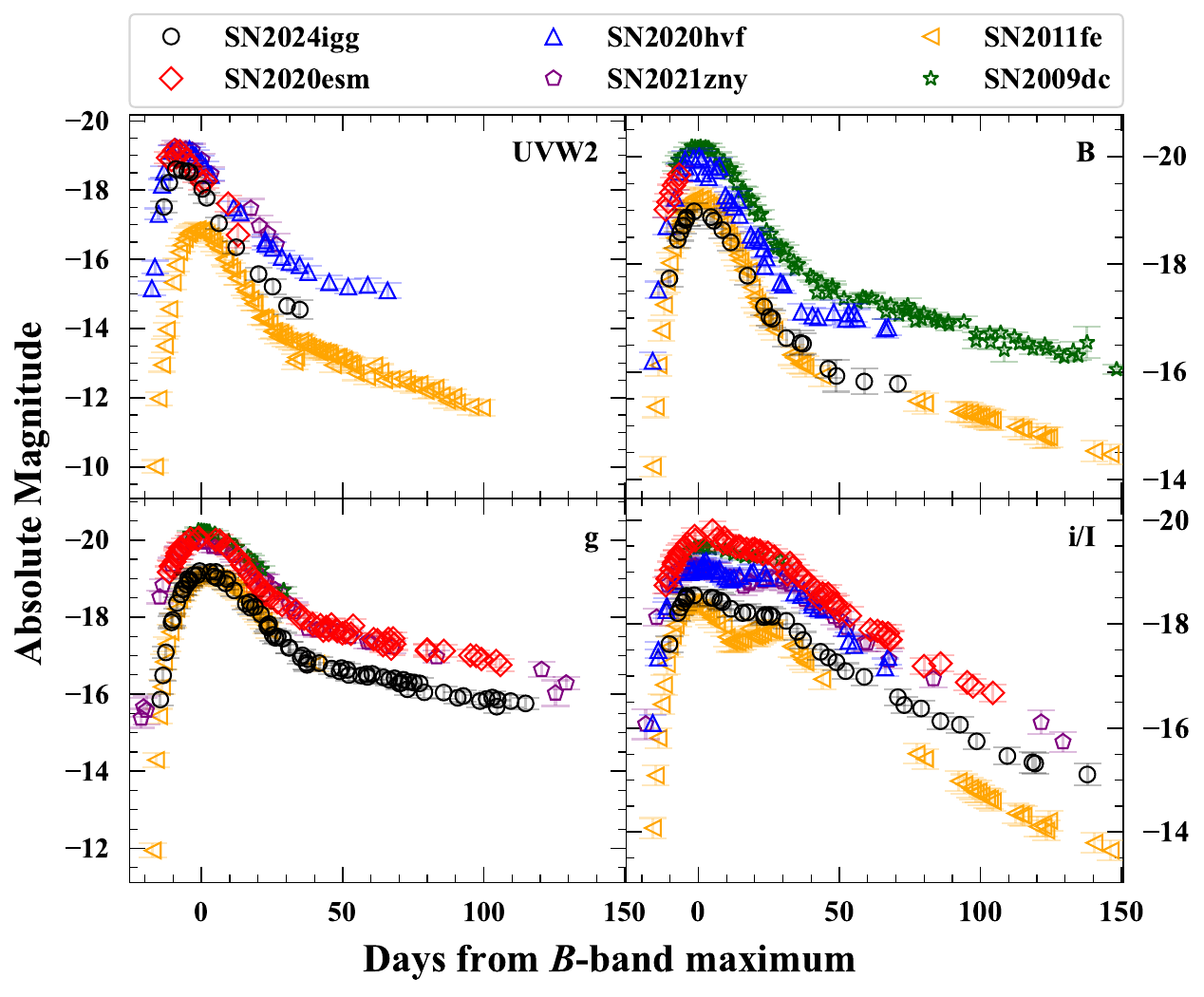}
        \includegraphics[width=\hsize]{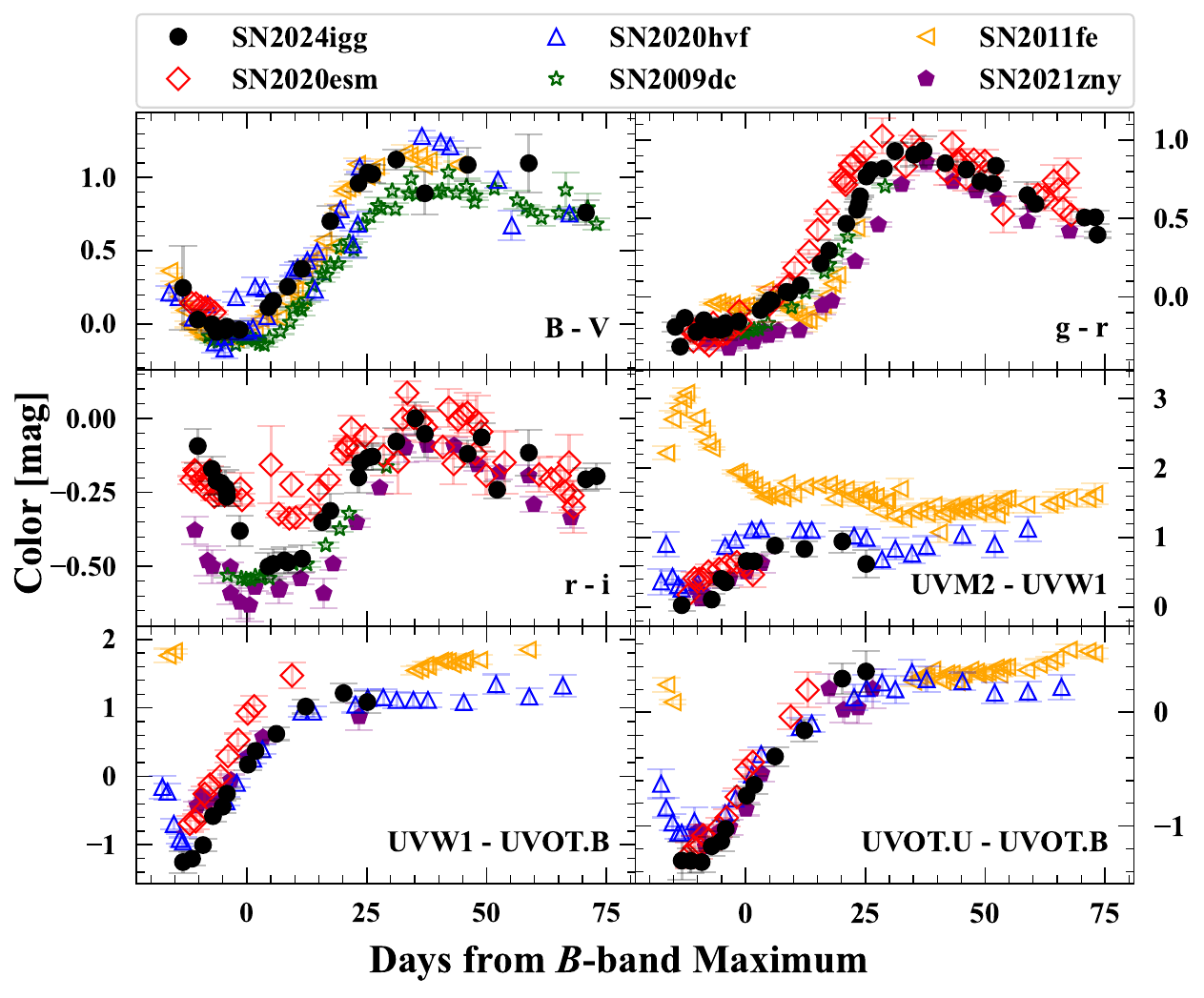}
        \caption{Light-curve and color comparisons of SN~2024igg with a typical normal Type Ia SN~2011fe and four 03fg-like objects: SNe~2009dc, 2020esm, 2020hvf, and 2021zny. All  light curves and colors have been corrected for total reddening. The light curves are shown in absolute magnitude. The $I$-band light curves of SN~2020hvf and 2011fe have been transformed to AB magnitude \citep{1983ApJ...266..713O}.}
        \label{fig:LC_compare}%
    \end{figure}

\begin{figure}
        \centering
        \includegraphics[width=\hsize]{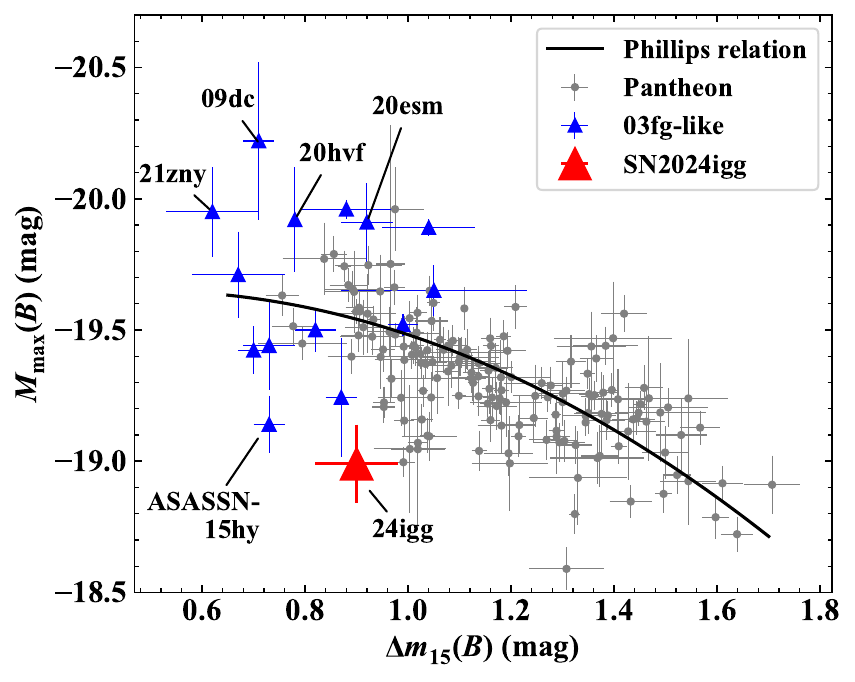}
        \caption{Comparison of the $B$-band light-curve decline rate and the absolute maximum magnitude for a sample of SNe Ia. SN~2024igg is emphasized as a red triangle. Normal SNe Ia from the Pantheon samples \citep{2018ApJ...859..101S} are shown as gray dots. The black solid curve represents the best-fit Lira-Phillips relation. The 03fg-like sample, shown as blue triangles, is taken from \citet{2021ApJ...922..205A} except for SNe~2009dc \citep{2011MNRAS.412.2735T}, 2020esm \citep{2022ApJ...927...78D}, 2020hvf \citep{2021ApJ...923L...8J}, and 2021zny \citep{2023MNRAS.521.1162D}. 
        }
        \label{fig:LWR}%
    \end{figure}

In Fig.~\ref{fig:LC_compare}, we compare the UV and optical light curves and colors of SN~2024igg with those of several well-observed SNe~Ia, including the normal SN~2011fe \citep{2015MNRAS.446.3895F,2016ApJ...820...67Z,2017MNRAS.472.3437G} and four 03fg-like objects: SNe~2009dc \citep{2010AJ....139..519C,2011MNRAS.412.2735T,2012MNRAS.425.1789S}, 2020esm \citep{2022ApJ...927...78D}, 2020hvf \citep{2021ApJ...923L...8J}, and 2021zny \citep{2023MNRAS.521.1162D}. The first two data points of $B-V$ color of SN~2024igg were calculated using $UVOT.B$ and $UVOT.V$, since they are earlier than the ground-based observations and the response curves are similar. SN~2024igg has a relatively low luminosity in the optical bands among 03fg-like SNe~Ia, $\sim$1 mag fainter than SNe~2020esm, 2020hvf and 2009dc. Actually, SN~2024igg exhibits a lower $B$-band luminosity than SN~2011fe initially, although this trend reverses at $t\gtrsim+25$ days after $B$-band maximum because of the slower decline ($\Delta m_{15}(B)=0.90\pm0.08$) rate compared to SN~2011fe \citep[$\Delta m_{15}(B)=1.18\pm0.03$;][]{2016ApJ...820...67Z}. 
We analyzed the position of SN~2024igg in the $B$-band WLR in Fig.~\ref{fig:LWR} and found it below the Lira-Phillips relation of normal SNe~Ia. 
Despite the relatively low optical luminosity, SN~2024igg shows remarkable similarities to other 03fg-like SNe in terms of (1) high UV luminosity ($UVW2<-$18 mag at $t\approx -10$ days), (2) small decline rate ($\Delta m_{15}(B)<1.0$ mag), and (3) absence of a prominent $i$-band secondary maximum that is clearly seen in normal SN~2011fe. Also, note that 03fg-like SNe with relatively low optical luminosities have already been found before, such as ASASSN-15hy \citep[$M_B\approx-19.14$ mag;][]{2021ApJ...920..107L} and SN~2022pul \citep[$M_B\approx-18.9$ mag;][]{2024ApJ...960...88S}.
 
The $B-V$ color of SN~2024igg evolves blueward before the $B$-band maximum, similar to normal SN~2011fe. But for $UVM2-UVW1$, SN~2024igg and SN~2011fe show inverse evolution trends at $t\gtrsim-10$ d: SN~2024igg reddens, while SN~2011fe continues blueward.  Consistent with the findings of \citet{2024ApJ...966..139H} for 03fg-like SNe, SN~2024igg is separated from SN~2011fe by $UVM2-UVW1 \gtrsim1.0$ mag at $t\approx\ $$-$10 days relative to $B$-band maximum. 
Together with the $g-r$ color that continuously reddens from peak up to about +30 days, and the decline of the UV flux proportion indicated by $UVOT.U-UVOT.B$ and $UVW1-UVOT.B$ colors even before the peak luminosity, these characteristics of SN~2024igg again resemble other 03fg-like objects.
We notice a linear evolution of the $r-i$ color of SN~2024igg before $B$-band maximum. This evolution trend is similar to that of SN~2021zny but the color is redder, which might be due to the strong C\,{\sc ii} $\lambda6580$ absorption of SN~2024igg that suppresses the $r$-band flux.

\subsection{Spectral analysis}\label{subsec:Spec}

    \begin{figure*}
        \centering
        \includegraphics[width=0.49\textwidth]{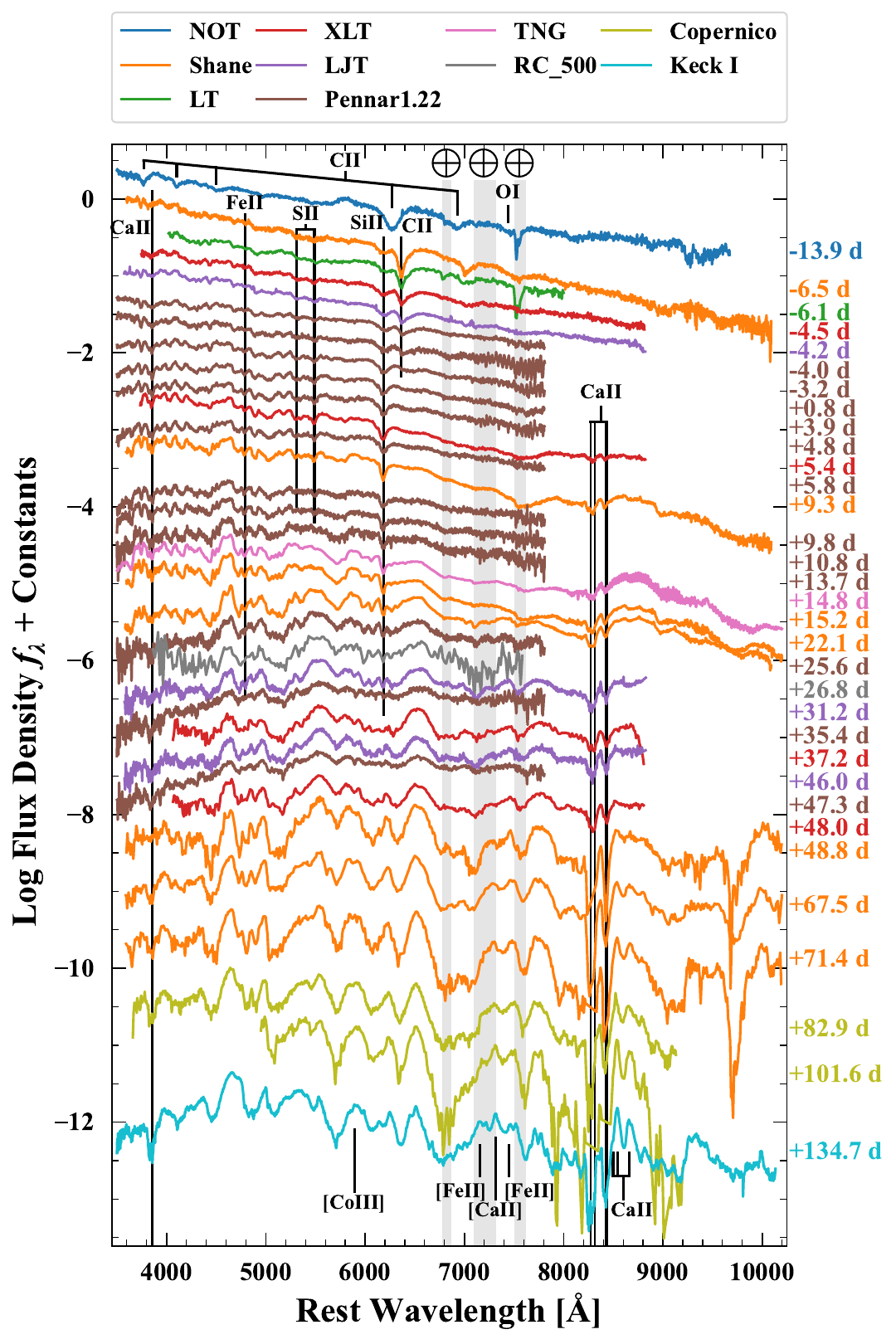}
        \includegraphics[width=0.49\textwidth]{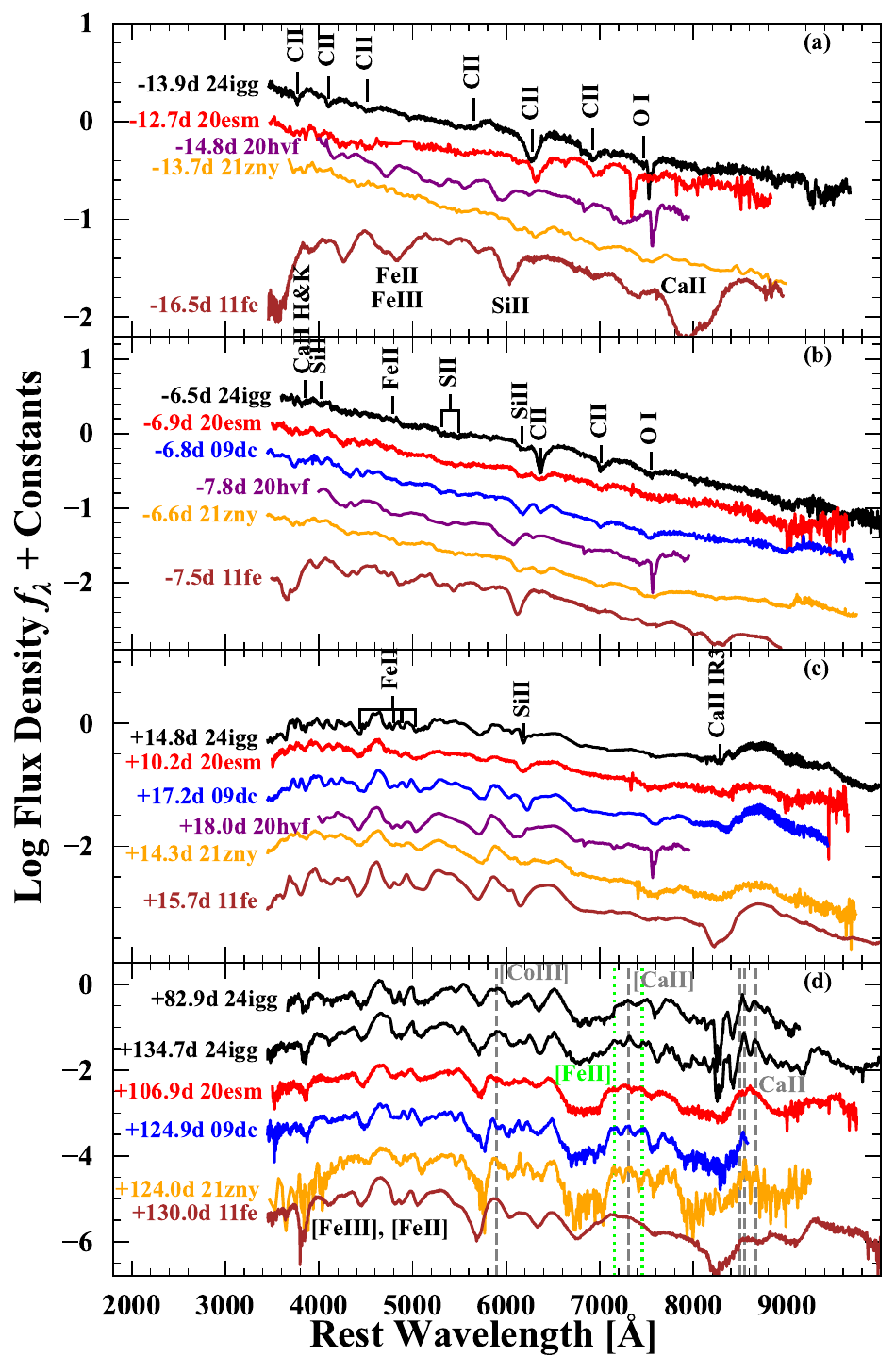}
        \caption{Optical spectral evolution of SN~2024igg and comparison with other SNe~Ia. All spectra have been corrected for reddening and host-galaxy redshift. In the left panel, we present the spectral evolution of SN~2024igg from $-13.9$ to $+$134.7 days relative to the $B$-band maximum. Spectra taken with different telescopes are in different colors as indicated in the top legend, and the phase of each spectrum is shown on the right side. Regions of the main telluric absorption are marked by gray vertical bands. The long solid lines indicate Ca\,{\sc II} H\&K, Fe\,{\sc II} $\lambda4924$, S\,{\sc II} $\lambda\lambda5454$, 5640, Si\,{\sc II} $\lambda6355$, and Ca\,{\sc II} $\lambda\lambda\lambda$8498, 8542, 8662  at $-$8000 km s$^{-1}$, as well as C\,{\sc II} $\lambda6580$ at $-$10,000 km s$^{-1}$. In the right panel, we show the spectral comparison of SN~2024igg and SNe~2009dc, 2011fe, 2020esm, 2020hvf, and 2021zny at selected epochs ($t \approx\ $$-$15, $-$7, $+$15, and $+$100 days). Spectral features are labeled with short lines in subplots (a), (b), and (c). In subplot (d), rest wavelengths of [Co\,{\sc II}] $\lambda5893$, [Ca\,{\sc II}] $\lambda7308$, and Ca\,{\sc II} $\lambda\lambda\lambda$8498, 8542, 8662 are indicated with gray dashed lines, and those of [Fe\,{\sc II}] $\lambda\lambda$7155, 7453 are indicated in lime dotted lines.} 
        \label{fig:spec_all}%
    \end{figure*}

To display the spectroscopic evolution of SN~2024igg, we present the entire spectral series with labels of some important lines in the left panel of Fig.~\ref{fig:spec_all}, together with spectral comparisons with other well-observed SNe at different phases in the right panel. The compared SNe include a normal Type Ia SN~2011fe \citep{2016ApJ...820...67Z, 2020MNRAS.492.4325S}, and four 03fg-like objects: SNe~2009dc \citep{2011MNRAS.410..585S, 2011MNRAS.412.2735T}, 2020esm \citep{2020TNSCR.861....1T, 2022ApJ...927...78D}, 2020hvf \citep{2023ApJ...943L..20S}, and 2021zny \citep{2023MNRAS.521.1162D}.

\subsubsection{First spectrum at -13.9 days}\label{subsubsec:frist_spec}

The first spectrum of SN~2024igg, taken at $t\approx\ $$-$13.9 days relative to the $B$-band maximum, shows a blue continuum and weak spectral features except for the prominent one at $\sim$6300 \AA\ which has a pseudo-equivalent width (pEW) of $\sim$90 \AA. These characteristics resemble the spectrum of SN~2020esm at $-12.6$ days, in which the $\sim$6300 \AA\ feature was attributed to C\,{\sc ii} $\lambda6580$. We ran the radiative transfer code \textsc{TARDIS} \citep{2014MNRAS.440..387K} to confirm this identification based on a toy model, in which blackbody radiation at the photoshpere with a velocity of 12,800 km s$^{-1}$ was assumed and the outer ejecta, with a total mass of $\sim$0.07 $M_{\rm \odot}$, were composed solely of carbon and oxygen in equal masses. An exponential density profile with an $e$-fold factor of 2000 km s$^{-1}$ was adopted. Local thermodynamic equilibrium (LTE) and dilute-LTE were assumed for ionization and excitation, respectively. An explosion time of $t_{\rm exp}=-18.4$ days, which is close to the time of the last nondetection and consistent with the fitting result in Section\,\ref{subsec:bolo}, was used to calculate the expansion time of the ejecta\footnote{More details of our \textsc{TARDIS} setup and \textsc{PYTHON} codes are available in \textsc{GITHUB} \url{https://github.com/jl-liu2022/TARDIS_SN2024igg}}. 

\begin{figure*}
    \centering
    \includegraphics[width=0.65\textwidth]{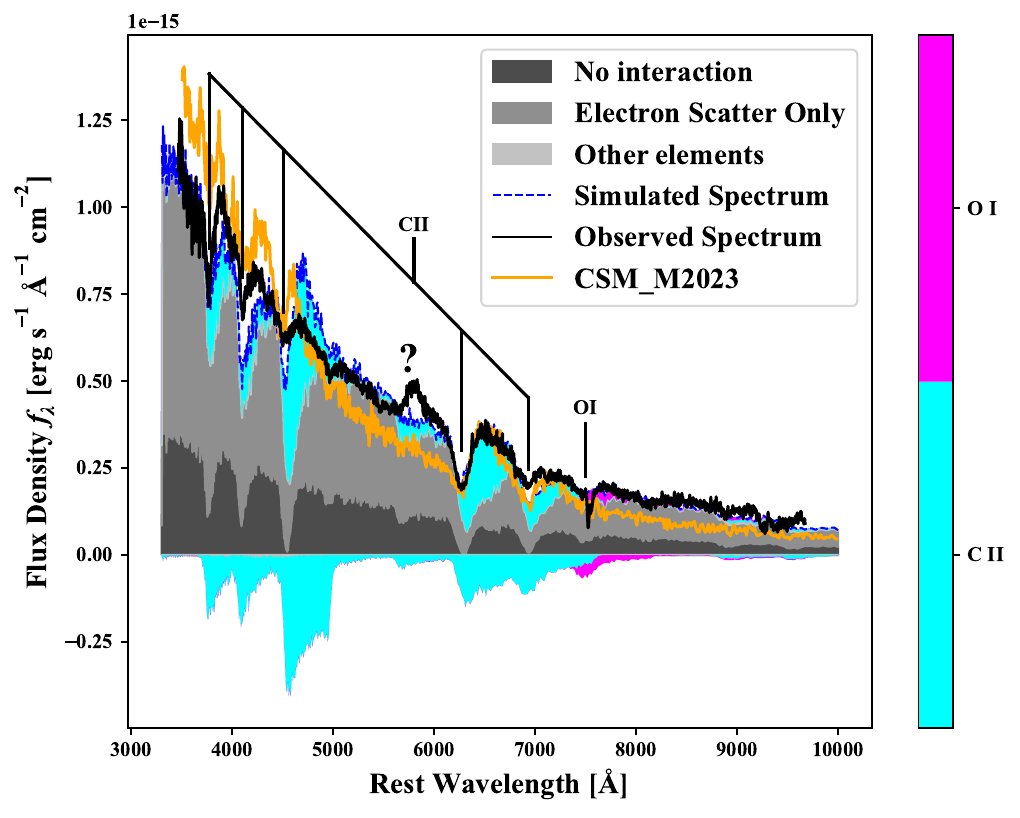}
    \caption{Comparison of the spectrum of SN~2024igg (black solid) at $-13.9$ days relative to $B$-band maximum and the synthetic spectrum using \textsc{TARDIS} (blue dashed) as well as a spectrum taken from the carbon-rich CSM interaction model (orange solid, labeled as CSM\_M2023) \citep{2023MNRAS.521.1897M}. The spectrum of SN~2024igg has been calibrated by the photometry and corrected for reddening and host redshift. The spectrum of CSM\_M2023, with a CSM mass of 0.1 $M_{\rm \odot}$, is shifted according to the emission peak at $\sim$6500 \AA\ for better display. The decomposition of the \textsc{TARDIS} spectrum is also shown in the plot, with the positive emission and negative absorption components labeled as different colors. The absorption features of C\,{\sc ii} and O\,{\sc i} are denoted by vertical lines.}
    \label{fig:spec_model}%
\end{figure*}

The simulated spectrum obtained with \textsc{TARDIS} is shown in Fig.~\ref{fig:spec_model}, along with a C/O-rich CSM interaction model spectrum at $t \approx\ $$-$16.6 days from \citet{2023MNRAS.521.1897M}. The overall flux of the \textsc{TARDIS} spectrum agrees well with the observed spectrum, indicating good estimates of the distance, extinction, and explosion time of SN~2024igg in this work. The strong C\,{\sc II} $\lambda6580$ feature is well reproduced by our configurations, but other carbon line strengths at $\sim$4100 and $\sim$4510 \AA\ are overestimated, which complicates the accurate estimation of the carbon mass. Nevertheless, most of the absorption features in the first spectrum of SN~2024igg could be identified as C\,{\sc ii}. This is consistent with the expectation of the CSM interaction model, in which a thermonuclear explosion occurs within a dense carbon-rich CSM with a mass of 0.1 $M_{\rm \odot}$. The usual lines seen in normal SNe~Ia (e.g., Fe\,{\sc ii} at $<$5000 \AA, S\,{\sc ii} $\lambda6355$, and ``W''-shaped S\,{\sc ii}) are still hidden beneath the photosphere at this phase.

We note that some features are not reproduced by either our toy model or the CSM model spectrum, especially the P-Cygni profile with a clear emission at $\sim$5800 \AA\ and a broad absorption blended with other lines at $\sim$5400--5700 \AA, both of which are not seen in other 03fg-like SNe. This mismatch might stem from some simplifications in the simulations like the LTE assumption, and the emission at $\sim$5800 \AA\ could still be attributed to a somehow blueshifted C\,{\sc ii} $\lambda5890$, as the spectrum is C\,{\sc ii}-dominated. We note that this feature also resembles the blueshifted He\,{\sc i} $\lambda5876$ of SN~2019odp \citep{2025A&A...693A..13S}. Other He lines, such as He\,{\sc i} $\lambda6678$, are not seen, possibly owing to their weakness or blending with C\,{\sc ii}. The origins of the weak absorption at $\sim$5470 \AA\ and $\sim$5570 \AA\ are also unclear; the bluer one may be attributed to Si\,{\sc iii} $\lambda5740$ with a blueshifted velocity of about $-14,000$ km s$^{-1}$.

\subsubsection{Spectral evolution}\label{subsubsec:following_spec}

The second spectrum of SN~2024igg was obtained at $t\approx\ $$-$6.5 days. At this epoch, weak absorption of the usual elements begins to appear with a velocity of about $-8000$ km s$^{-1}$ (measured from the local minimum), such as Fe\,{\sc ii} $\lambda4924$, ``W''-shaped S\,{\sc ii} absorption trough, Si\,{\sc ii} $\lambda4130$ and $\lambda6355$, and Ca\,{\sc ii} H\&K (contaminated by C\,{\sc ii} $\lambda3921$). These weak features at low velocities, alongside the prominent C\,{\sc ii} $\lambda6580$ absorption, resemble those of other 03fg-like SNe except for SN~2020hvf. Notably, the C\,{\sc ii} $\lambda6580$ line of SN~2024igg has some unique properties at this epoch. In other 03fg-like SNe, C\,{\sc ii} $\lambda6580$ has a strength comparable to or less than Si\,{\sc ii} $\lambda6355$. In SN~2024igg, however, it remains the strongest absorption feature.  
Moreover, it is overwhelmingly strong (pEW $\approx66$ \AA) compared to any other 03fg-like SNe \citep[pEW $\lesssim30$ \AA;][]{2021ApJ...922..205A} at this epoch. \citet{2021ApJ...922..205A} found a negative correlation between the Si\,{\sc ii} velocity and the strength of C\,{\sc ii} among 03fg-like SNe, which also holds for SN~2024igg, as the Si\,{\sc ii} $\lambda6355$ velocity of $-8000$ km s$^{-1}$ is also among the lowest ever recorded for this subtype of SN~Ia. 
The C\,{\sc ii} $\lambda6580$ absorption is likely detached above the photosphere, as it has a sharp profile and a higher velocity (about $-10,000$ km s$^{-1}$) compared to Si\,{\sc ii}. Given that the line opacity scales with the density, this line profile implies that the carbon density peaks around $-$10,000 km s$^{-1}$ and drops rapidly toward the inner region.

Examining the left panel of Fig.~\ref{fig:spec_all}, we see that almost all the line velocities evolve slowly before vanishing, suggesting that line-forming regions are narrow. An exception is C\,{\sc ii} $\lambda6580$ that suddenly broadens to a flat profile and drops to $\sim$ $-$7000 km s$^{-1}$ around peak, possibly due to the existence of another component of carbon in the inner layer.

As SN~2024igg evolves at $t\approx\ $$+$14.8 days, Si\,{\sc ii} $\lambda6355$ weakens, C\,{\sc ii} $\lambda6580$ and the ``W''-shaped S\,{\sc ii} disappear, and the Fe\,{\sc ii} lines become more prominent. While the spectrum of SN~2011fe shows a strong and blended Ca\,{\sc ii} near-infrared (NIR) triplet, the spectrum of SN~2024igg  exhibits a relatively weak but separated Ca\,{\sc ii} $\lambda8662$ and a blend of Ca\,{\sc ii} $\lambda\lambda8498$, 8542, bearing a strong resemblance to that of SN~2009dc. By $t\approx $$+$100 days, SN~2024igg enters the early nebular phase when the outer ejecta become moderately transparent. Forbidden lines like [Fe\,{\sc ii}], [Fe\,{\sc iii}], and [Co\,{\sc iii}] are seen in  spectra of both 03fg-like and normal SNe~Ia at this phase. A key distinction of SN~2024igg from SN~2011fe is the presence of resolved lines, especially at redder wavelengths. This characteristic, combined with its centrally peaked Ca\,{\sc ii}, is again akin to SNe~2009dc and 2021zny. In contrast, the lines in SN 2020esm appear broader and its Ca\,{\sc ii} lines are blueshifted. 

More resolved lines appear in the spectrum taken at $t\approx\ $$+$134.7 days, in which we identified forbidden emissions from [Co\,{\sc iii}] $\lambda5893$ (weighted average of the doublet), [Ca\,{\sc ii}] $\lambda7313$ (weighted average of the doublet), and [Fe\,{\sc ii}] $\lambda\lambda7155$, 7453, all peaking at their rest wavelengths. Other 03fg-like SNe also show [Ca\,{\sc ii}], but they do not reveal the resolved [Co\,{\sc iii}] and [Fe\,{\sc ii}] at rest.
Although a narrow line is also present near 7155 \AA\ in SN~2021zny, it might not be attributed to [Fe\,{\sc ii}] $\lambda7155$. The reason is that the spectrum has a low signal-to-noise ratio and that the corresponding [Fe\,{\sc ii}] $\lambda7453$ line (sharing the same upper level) is not visible. 

\subsection{Pseudobolometric light curve}\label{subsec:bolo}

We constructed the pseudobolometric (1600--24,000 \AA) light curve of SN~2024igg using {\it Swift} $UVW2$, $UVM2$, $UVW1$, $U$, and ground-based $BgVri$ photometry. The UV and optical luminosities were calculated by directly integrating the photometry flux densities. To estimate the NIR luminosity of SN~2024igg, we assumed a similar ratio of the NIR to the optical luminosity to that of SN~2009dc, which shows similar color evolution in $r-i$ (after peak) and $g-r$. But we caution that SN~2009dc is much more luminous than SN~2024igg, and the NIR flux proportion varies significantly among 03fg-like SNe \citep{2021ApJ...922..205A}. More details of the calculations can be found in Appendix\,\ref{sec:bolo_table}. 

\begin{figure}
        \centering
        \includegraphics[width=\hsize]{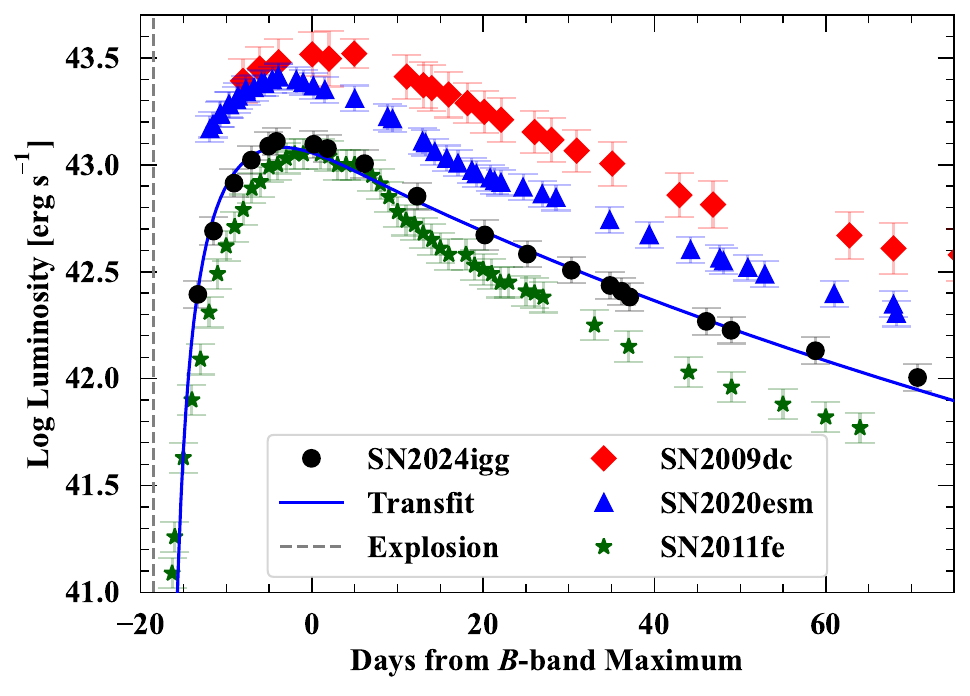}
        \caption{Comparison of the bolometric light curves of SN 2024igg (black circles) with those of SNe 2009dc (blue triangles), 2011fe (green stars), and 2020esm (red diamonds). The best-fit model from \textsc{Transfit} for SN 2024igg is overplotted as a blue line, and the corresponding explosion time is indicated by the gray dashed line.}
        \label{fig:bolo}%
    \end{figure}

The pseudobolometric light curve of SN~2024igg from $-13.3$ to $+70.7$ days is shown in Fig.~\ref{fig:bolo}, in which we also include those of SNe~2009dc, 2011fe \citep{2016ApJ...820...67Z}, and 2020esm. The bolometric light curve of SN~2009dc was recalculated using the photometry from several sources \citep{2010AJ....139..519C,2011MNRAS.412.2735T,2012MNRAS.425.1789S,2015ApJS..220....9F}. 
As expected from the light-curve comparisons in Section~\ref{subsec:LC_evolution}, SN~2024igg has a relatively low bolometric luminosity compared to other 03fg-like objects such as SNe~2009dc and 2020esm, but still more luminous than SN~2011fe due to its high UV luminosity. The decline rate of the pseudobolometric light curve of SN~2024igg is stable, similar to those of the 03fg-like but different from that of SN~2011fe, which shows a relatively fast decline during the first $\sim$15 days after peak brightness. We performed a fourth-order polynomial fit to the pseudobolometric light curve and obtained a peak time of $t_{\rm peak,bol}=60450.75\pm0.39$ (MJD), which is earlier than the $B$-band maximum by $\sim$1.34 days, and a peak luminosity of $L_{\rm peak,bol}= (1.31\pm0.18) \times10^{43}$ erg s$^{-1}$. 

Assuming that the pseudobolometric luminosity of SN~2024igg is only due to the radioactive decay of $^{56}$Ni, we used a numerical light-curve fitting tool \textsc{Transfit} \citep{2025ApJ...992...20L} to estimated the mass of $^{56}$Ni ($M_{\rm Ni}$) synthesized in the explosion and the total ejecta mass ($M_{\rm ej}$). Compared to the commonly used Arnett analytic model \citep{1982ApJ...253..785A}, \textsc{Transfit} solves the time-dependent energy balance and photon diffusion numerically, and thus does not rely on the restrictive assumptions of a single-zone and strictly homologous/instantaneous deposition treatment. This allows a more flexible and self-consistent mapping between $M_{\rm ej}$, $M_{\rm Ni}$, $E_{\rm K}$, and the light-curve shape, and is well suited for Bayesian inference with robust uncertainty quantification. 
The best fit is presented in Fig.~\ref{fig:bolo}, which corresponds to $M_{\rm ej}=1.54^{+0.22}_{-0.19}$ $M_{\rm \odot}$ and $M_{\rm Ni}=0.547^{+0.082}_{-0.082}$ $M_{\rm \odot}$ (distance error incorporated). More details of the fitting as well as the Markov Chain Monte Carlo (MCMC) corner plot can be found in Appendix\,\ref{sec:bolo_table}. The fitting also gives an explosion time relative to $B$-band maximum of $t_{\rm exp}=-18.29^{+0.27}_{-0.14}$ days, which is $\sim$1.5 days before the time of first light estimated by the power-law fit in Section\,\ref{subsec:LC_evolution}. This supports a delay between the explosion and the first light powered by $^{56}$Ni. 

We note that $M_{\rm ej}$ is strongly correlated with another fitting parameter, the scaled velocity $v_{\rm sc}$ \citep{1982ApJ...253..785A}. If we estimate $v_{\rm sc}$ using the Si velocity at peak ($\sim$8000 km s$^{-1}$) instead of the fitting one ($v_{\rm sc}=10,350^{+930}_{-840}$ km s$^{-1}$) as some studies did (e.g., \citealt{2024PASP..136i4201B}), $M_{\rm ej}$ would drop to only $\sim$1 $M_{\rm \odot}$, below $M_{\rm Ch}$. For a validation of the fitting results, we also use the \textsc{PYBOLOSN} code \citep{2014MNRAS.440.1498S} to estimate $M_{\rm ej}$, $M_{\rm Ni}$, and $v_{\rm sc}$ of SN~2024igg. Assuming a rise time to the peak bolometric luminosity of $t_{\rm rise,bol}=15.65\pm1.50$ days (limited by the last nondetection and the first detection in ZTF $r$), we obtain $M_{\rm ej}=1.48^{+0.19}_{-0.12}$ $M_{\rm \odot}$, $M_{\rm Ni}=0.54^{+0.18}_{-0.14}$ $M_{\rm \odot}$, and $v_{\rm sc}=10,130^{+710}_{-610}$ km s$^{-1}$, all of which are consistent with the \textsc{Transfit} results within the uncertainties. 

\section{Discussion}\label{sec:Discussions}

The photometric and spectroscopic observations of SN~2024igg, such as the slowly declining light curve, blue UV$-$optical colors, and strong carbon features, indicate that this SN is a member of the 03fg-like subclass of SNe Ia. These SNe~Ia are thought to explode inside a carbon-rich CSM given their prominent and long-lived C\,{\sc ii} features (e.g., \citealt{2021ApJ...920..107L,2021ApJ...922..205A,2023MNRAS.521.1162D}), especially for SN~2020esm whose early-time spectra are dominated by carbon and oxygen absorption \citep{2022ApJ...927...78D}. The C\,{\sc ii}-dominated first spectrum of SN~2024igg, with even stronger C\,{\sc ii} line strengths and absence of Si\,{\sc ii} compared to SN~2020esm, further supports the carbon-rich CSM interaction model \citep{2023MNRAS.521.1897M}. Specifically, the interaction with the CSM leads to the formation of a photosphere within the swept-up, carbon-rich envelope. This outer photosphere effectively masks the elemental signatures of the underlying SN ejecta; thus, only features from the CSM itself are detectable at very early times. These features are more prominent in the first spectrum of SN~2024igg possibly due to its observation at an even earlier phase than that of SN~2020esm. 

Other observational properties of SN~2024igg also find a unified explanation in the context of CSM interaction. The luminous UV excess, relative to normal SNe Ia, may be interpreted as a reduction in line blanketing due to the hiding of iron-group elements. Concurrently, the mass gain from the interaction consistently accounts for the marginally super-Chandrasekhar mass and the slowly declining light curve. Furthermore, as the ejecta are decelerated and compressed by the CSM \citep{2016MNRAS.463.2972N}, the spatial extent of line formation is reduced, leading to the observed weak absorption lines with low and slowly evolving velocities \citep{2005A&A...437..667D,2005A&A...439..671D,2014MNRAS.441..532D}. Based on the persistence of line velocities near $-$10,000 km s$^{-1}$ for C\,{\sc ii} $\lambda6580$ and $-$8000 km s$^{-1}$ for lines of intermediate-mass elements (IMEs), we speculate that the majority of the SN ejecta were compressed to $\lesssim-$10,000 km s$^{-1}$ and the density peaked at about $-8000$ km s$^{-1}$. 

The carbon-rich CSM of 03fg-like SNe could stem from the classical double degenerate (DD) scenario \citep{1984ApJ...277..355W}, 
in which a thermonuclear explosion is triggered on a secular timescale (much longer than a dynamical timescale) after the onset of a WD-WD merger.
The secondary C/O WD is totally disrupted, forming a C/O Keplerian disk \citep{2007MNRAS.380..933Y} that subsequently accretes onto the primary. 
Though the primary WD could evolve to an ONeMg WD that subsequently collapses to a neutron star in this context \citep{1985A&A...150L..21S, 1998ApJ...500..388S}, 
this could be avoided in some cases and a thermonuclear explosion might be eventually triggered if the WD mass reaches the $M_{\rm Ch}$ limit \citep{2007MNRAS.380..933Y}.
Different from the slow evolution of the classical scenario, the thermonuclear explosion could also be triggered during the dynamical merger of the WD binary \citep[dubbed as ``violent merger'';][]{2012ApJ...747L..10P}. Though the ejected material during the dynamical merger could be confined owing to the short timescale in this scenario, a wind driven by super-Eddington accretion before the dynamical merger could form an extended CSM \citep{2025arXiv251210014I}. Also, note that an aspherical explosion of the violent merger is supported by the high intrinsic polarization observed in 03fg-like SNe~2021zny and 2022ilv \citep{2024A&A...687L..19N}. 
For SN~2024igg, however, the absence of clear shifts in its nebular-phase forbidden lines disfavors an aspherical explosion. 

In the core-degenerate (CD) scenario \citep{2011MNRAS.417.1466K}, a massive WD produced by the merger of a WD with the degenerate core of an AGB star may explode as an SN~Ia. In this picture, if the envelope of the AGB star is not completely ejected when the explosion occurs, it would serve as a carbon-rich CSM and interact with the SN ejecta \citep{2020ApJ...900..140H,2021ApJ...920..107L,2021ApJ...922..205A}. As the essence of the CD scenario is also a double-WD merger, the main observational difference from the classical double-degenerate (DD) scenario should stem from the AGB star envelope that might be H-rich and/or He-rich. A line possibly associated with blueshifted He\,{\sc I} $\lambda5876$ was found in the earliest spectrum of SN~2024igg, but an attribution to C\,{\sc I} $\lambda5890$ is also possible. Flash features of the spectra within the interaction phase may help identify this line and distinguish the DD and CD scenarios. 

As the luminosity is moderate and the inferred ejecta mass marginally exceeds $M_{\rm Ch}$, this SN may come from a $M_{\rm Ch}$ explosion, such as a delayed-detonation (e.g., \citealt{2005ApJ...623..337G,2013MNRAS.429.1156S}), within a dense carbon-rich CSM with a mass of $\sim$0.1 $M_{\rm \odot}$. The [C\,{\sc I}] and [O\,{\sc I}] lines, which are powerful for constraining the explosion mechanism (e.g., \citealt{2025ApJ...982L..18L}), are not detected in SN~2024igg, but this might be due to the absence of observations at sufficiently late epochs.     

\section{Conclusions}\label{sec:Conclusions}

In this work, we presented and analyzed photometric and spectroscopic observations of SN~2024igg. Though it has a relatively low optical luminosity  ($M_{\rm max}(B)=-18.99\pm0.15$ mag) and short rise time (less than 18.5 days to $B$-band maximum), 
it shares many characteristics with other 03fg-like SNe, including the high UV flux, blue UV$-$optical colors, slowly declining light curve ($\Delta m_{15}(B)=0.90\pm0.08$ mag), lack of a prominent $i$-band secondary maximum, strong and long-lived C\,{\sc ii}, and weak spectral features of IMEs and iron-group elements at early times. Meanwhile, this SN exhibits the strongest C\,{\sc ii} $\lambda6580$ ever detected in an 03fg-like SN, with the pEWs of $\sim$90 \AA\ and $\sim$66 \AA\ at $t\approx\ $$-$13.9 and $-$6.5 days relative to $B$-band maximum, respectively. Remarkably, line velocities of IMEs are already as low as about $-8000$ km s$^{-1}$ at $t\approx\ $$-$6.5 days and evolve slowly until the lines disappear. Together with the C\,{\sc ii}-dominated spectrum at $t\approx\ $$-$13.9 days, these properties of SN~2024igg strongly support the idea that this SN exploded within a carbon-rich CSM, which could originate from the debris of the donor WD in a double-WD merger or the envelope of an AGB star. As the forbidden lines in the early nebular-phase spectrum are unshifted, we suggest that this SN originates from a symmetric explosion that is triggered on a secular timescale after the merging occurred.



\begin{acknowledgements}

This work is supported by the National Natural Science Foundation of China (NSFC grants 12288102, 12033003, 11633002) and the Tencent Xplorer Prize. We thank Yi Yang for his contributions to the Keck proposal development and for assisting with the SN 2024igg observations carried out by the Berkeley group.
A.V.F.'s group at UC Berkeley received financial assistance from the Christopher R. Redlich Fund, as well as donations from Gary and Cynthia Bengier, Clark and Sharon Winslow, Alan Eustace and Kathy Kwan, William Draper, Timothy and Melissa Draper, Briggs and Kathleen Wood, Ellen and Alan Seelenfreund (W.Z. is a Bengier-Winslow-Eustace Specialist in Astronomy, T.G.B. is a Draper-Wood-Seelenfreund Specialist in Astronomy), 
and numerous other donors.
Y.-Z. Cai is supported by NSFC grant 12303054, the National Key Research and Development Program of China (grant  2024YFA1611603), and the Yunnan Fundamental Research Projects (grants  202401AU070063, 202501AS070078).
A.P., A.R., and G.V. acknowledge support from the PRIN-INAF 2022, ``Shedding light on the nature of gap transients: from the observations to the models.'' A.R. was also supported by the GRAWITA Large Program Grant (PI P. D'Avanzo).
J.-J. Zhang and Y.-Z. Cai are supported by the International Centre of Supernovae, Yunnan Key Laboratory (grant 202302AN360001). J.-J. Zhang is supported by the National Key R\&D Program of China  (grant 2021YFA1600404), NSFC grant 12173082, the Yunnan Province Foundation (grant 202201AT070069), the Top-notch Young Talents Program of Yunnan Province, and the Light of West China Program provided by the Chinese Academy of Sciences. \\

\indent We thank the staff at the various observatories where data were obtained.
A major upgrade of the Kast spectrograph on the Shane 3~m telescope at Lick Observatory, led by Brad Holden, was made possible through generous gifts from the Heising-Simons Foundation, William and Marina Kast, and the University of California Observatories. Research at Lick Observatory is partially supported by a generous gift from Google. 
Some of the data presented herein were obtained at the W. M. Keck Observatory, which is operated as a scientific partnership among the California Institute of Technology, the University of California, and NASA; the observatory was made possible by the generous financial support of the W. M. Keck Foundation.
This work is partially based on observations collected at Copernico and Schmidt telescope (Asiago, Mt. Ekar, Italy) of the INAF -- Osservatorio Astronomico di Padova, and at the Galileo telescope (Asiago, Mt. Pennar, Italy) of the Padova University.
This work is also based on observations made in the Observatorios de Canarias del IAC with the Telescopio Nazionale Galileo, operated on the island of La Palma by INAF at the Observatorio del Roque de los Muchachos under the program A50TAC\_41 (PI G. Valerin). \\

\indent This work was partially supported by the Open Project Program of the Key Laboratory of Optical Astronomy, National Astronomical Observatories, Chinese Academy of Sciences. 
Funding for the LJT has been provided by the CAS and the People's Government of Yunnan Province. The LJT is jointly operated and administrated by YNAO and Center for Astronomical Mega-Science, CAS.
ZTF is supported by the U.S. NSF under grants AST-1440341 and AST-2034437, and a collaboration including current partners Caltech, IPAC, the Oskar Klein Center at Stockholm University, the University of Maryland, the University of California (Berkeley), the University of Wisconsin at Milwaukee, University of Warwick, Ruhr University, Cornell University, Northwestern University, and Drexel University. Operations are conducted by COO, IPAC, and UW. \\

\indent This research made use of \textsc{tardis}, a community-developed software package for spectral
synthesis in supernovae \citep{2014MNRAS.440..387K, kerzendorf_2026_18210479}. The
development of \textsc{tardis} received support from GitHub, the Google Summer of Code
initiative, and from ESA's Summer of Code in Space program. \textsc{tardis} is a fiscally
sponsored project of NumFOCUS. \textsc{tardis} makes extensive use of Astropy and Pyne.
This work makes use of the NASA/IPAC Extragalactic Database, which is funded by NASA and operated by the California Institute of Technology.

\end{acknowledgements}

%

\bibliographystyle{aa}
\bibliography{aa.bib}

\begin{appendix}




\onecolumn
\section{Photometric and spectroscopic observations} \label{sec:observations_table}

The photometric data and the journal of spectroscopic observations are presented in Table\,\ref{table:lc} and Table\,\ref{table:spec}, respectively. The spectra will be available through the Weizmann Interactive Supernova Data Repository (WISeREP)\footnote{\url{https://www.wiserep.org/}}.

\begin{table*}[h]
\centering
\caption{Photometric Observations of SN\,2024igg.}\label{table:lc}
\begin{tabular}{cccccc}
\hline
MJD\tablefootmark{a} & Magnitude & Error\tablefootmark{b} & Band & Source \\
\hline
60436.451 & 19.101 & 0.078 & r & ZTF \\
60437.263 & 17.909 & 0.023 & r & ZTF \\
60437.304 & 17.730 & 0.017 & g & ZTF \\
60438.260 & 17.128 & 0.011 & r & ZTF \\
60438.282 & 17.409 & 0.024 & r & ZTF \\
60438.326 & 17.104 & 0.011 & g & ZTF \\
60438.636 & 15.675 & 0.062 & UVW1 & UVOT \\
... & ...  & ... & ... & ... \\
60591.437 & 18.602 & 0.148 & r & TNT \\
60591.441 & 18.466 & 0.150 & i & TNT \\
\hline
\end{tabular}

\tablefoot{\\
\tablefoottext{a}{This table is available in its entirety in machine-readable form.}\\
\tablefoottext{b}{$1\sigma$.}\\
}
\end{table*}

\begin{table*}[h]
\centering
\caption{\label{table:spec}Log of optical spectra of SN~2024igg.}
\begin{tabular}{lllllll}
\hline
\hline
MJD & Date & Phase\tablefootmark{a} & Range (\AA) & Exp. time (s) & Airmass & Instrument/Telescope \\
\hline
60438.0 & 20240507 & -13.9 & 3502-9777 & 1800 & 1.15 & ALFOSC/NOT \\
60445.5 & 20240515 & -6.5 & 3636-10200 & 900 & 1.42 & Kast/Shane \\
60445.9 & 20240515 & -6.1 & 4066-8086 & 900 & 1.22 & SPRAT/LT \\
60447.6 & 20240517 & -4.5 & 3787-8908 & 1800 & 1.08 & BFOSC/XLT \\
60447.9 & 20240517 & -4.2 & 3613-8918 & 2401 & 1.54 & YFOSC/LJT \\
60448.0 & 20240518 & -4.0 & 3315-7890 & 3600 & 1.09 & B\&C/Pennar1.22 \\
60448.9 & 20240518 & -3.2 & 3319-7894 & 1200 & 1.04 & B\&C/Pennar1.22 \\
60453.0 & 20240522 & +0.8 & 3316-7891 & 3600 & 1.01 & B\&C/Pennar1.22 \\
60456.0 & 20240526 & +3.9 & 3315-7888 & 3600 & 1.07 & B\&C/Pennar1.22 \\
60457.0 & 20240526 & +4.8 & 3318-7891 & 3600 & 1.02 & B\&C/Pennar1.22 \\
60457.6 & 20240527 & +5.4 & 3777-8917 & 2100 & 1.04 & BFOSC/XLT \\
60458.0 & 20240528 & +5.8 & 3315-7890 & 3600 & 1.09 & B\&C/Pennar1.22 \\
60461.5 & 20240531 & +9.3 & 3634-10200 & 900 & 1.62 & Kast/Shane \\
60462.0 & 20240531 & +9.8 & 3316-7893 & 5400 & 1.11 & B\&C/Pennar1.22 \\
60463.0 & 20240601 & +10.8 & 3318-7891 & 1800 & 1.09 & B\&C/Pennar1.22 \\
60465.9 & 20240604 & +13.7 & 3319-7892 & 1200 & 1.02 & B\&C/Pennar1.22 \\
60467.0 & 20240606 & +14.8 & 3342-10523 & 1200 & 1.16 & LRS/TNG \\
60467.5 & 20240606 & +15.2 & 3636-10200 & 1200 & 1.68 & Kast/Shane \\
60474.4 & 20240613 & +22.1 & 3624-10756 & 1500 & 1.33 & Kast/Shane \\
60478.0 & 20240616 & +25.6 & 3319-7891 & 3600 & 1.08 & B\&C/Pennar1.22 \\
60479.1 & 20240618 & +26.8 & 3907-7665 & 6000 & 1.06 & ATIK 460EX/RC\_500 \\
60483.6 & 20240622 & +31.2 & 3616-8923 & 2000 & 1.13 & YFOSC/LJT \\
60487.9 & 20240626 & +35.4 & 3317-7890 & 1800 & 1.10 & B\&C/Pennar1.22 \\
60489.7 & 20240628 & +37.2 & 4109-8902 & 3300 & 1.24 & BFOSC/XLT \\
60498.6 & 20240707 & +46.0 & 3616-8920 & 2000 & 1.22 & YFOSC/LJT \\
60499.9 & 20240708 & +47.3 & 3316-7889 & 9000 & 1.17 & B\&C/Pennar1.22 \\
60500.6 & 20240709 & +48.0 & 4109-8901 & 3600 & 1.13 & BFOSC/XLT \\
60501.4 & 20240710 & +48.8 & 3636-10726 & 600 & 1.60 & Kast/Shane \\
60520.3 & 20240729 & +67.5 & 3629-10744 & 1468 & 1.60 & Kast/Shane \\
60524.3 & 20240802 & +71.4 & 3632-10722 & 2100 & 1.54 & Kast/Shane \\
60535.9 & 20240813 & +82.9 & 3700-9230 & 5400 & 1.25 & AFOSC/Copernico \\
60554.8 & 20240901 & +101.6 & 5004-9296 & 1800 & 1.29 & AFOSC/Copernico \\
60588.2 & 20241005 & +134.7 & 3182-10244 & 450 & 2.09 & LRIS/Keck I \\ 
\hline
\end{tabular}
\tablefoot{\\
\tablefoottext{a}{Rest frame, with respect to the $B$-band maximum light, MJD$_{B\rm max}=60452.1$.}}\\
\end{table*}

\section{Construction and Fitting of pseudobolometric light curve} \label{sec:bolo_table}

The pseudobolometric light curve of SN~2024igg was constructed using {\it Swift} $UVW2$, $UVM2$, $UVW1$, $U$, and ground-based $BgVri$ photometry. {\it Swift} $UVOT.B$ and $UVOT.V$ were used at $t\approx\ $$-$13.31 and $-$11.48 days relative to $B$-band maximum, when the ground-based $BV$ photometry was not yet available. We assumed a linear evolution of the $r-i$ color, as found in Section~\ref{fig:LC_compare}, to estimate the $i$-band magnitude based on ZTF $r$. We assumed constant $UVW2-B$, $UVM2-B$, $UVW1-B$, and $UVOT.U-B$ colors at $t>+35$ days when UV photometry is not available, and note that the UV contribution to the total luminosity is small at this phase. The data were interpolated in each filter at the observed epochs of $UVM2$ ($t<+35$ days) and $B$ ($t>+35$ days) so that the pseudobolometric luminosity can be calculated at the same phase. After correcting for the Galactic reddening, the photometric magnitudes were converted to monochromatic fluxes. The spectral energy distribution (SED) was then interpolated linearly and integrated with respect to wavelength, assuming zero flux at the blue edge of the $UVW2$ band (1600 \AA). 

To estimate the flux of the 7500--24,000 \AA\ range, beyond the $i$-band, we assumed that the flux ratio of this range to the optical one (4400--7500 \AA) was similar to that of SN~2009dc, which showed a similar color evolution in $r-i$ (after peak) and $g-r$. Finally, a pseudobolometric (1600--24,000 \AA) light curve of SN~2024igg from $-13.31$ to $+70.72$ days was obtained by combining the UV, optical, and NIR luminosities. The  pseudobolometric light curve is presented in Table\,\ref{table:bolo_lc}.  

\begin{figure*}
        \centering
        \includegraphics[width=\hsize]{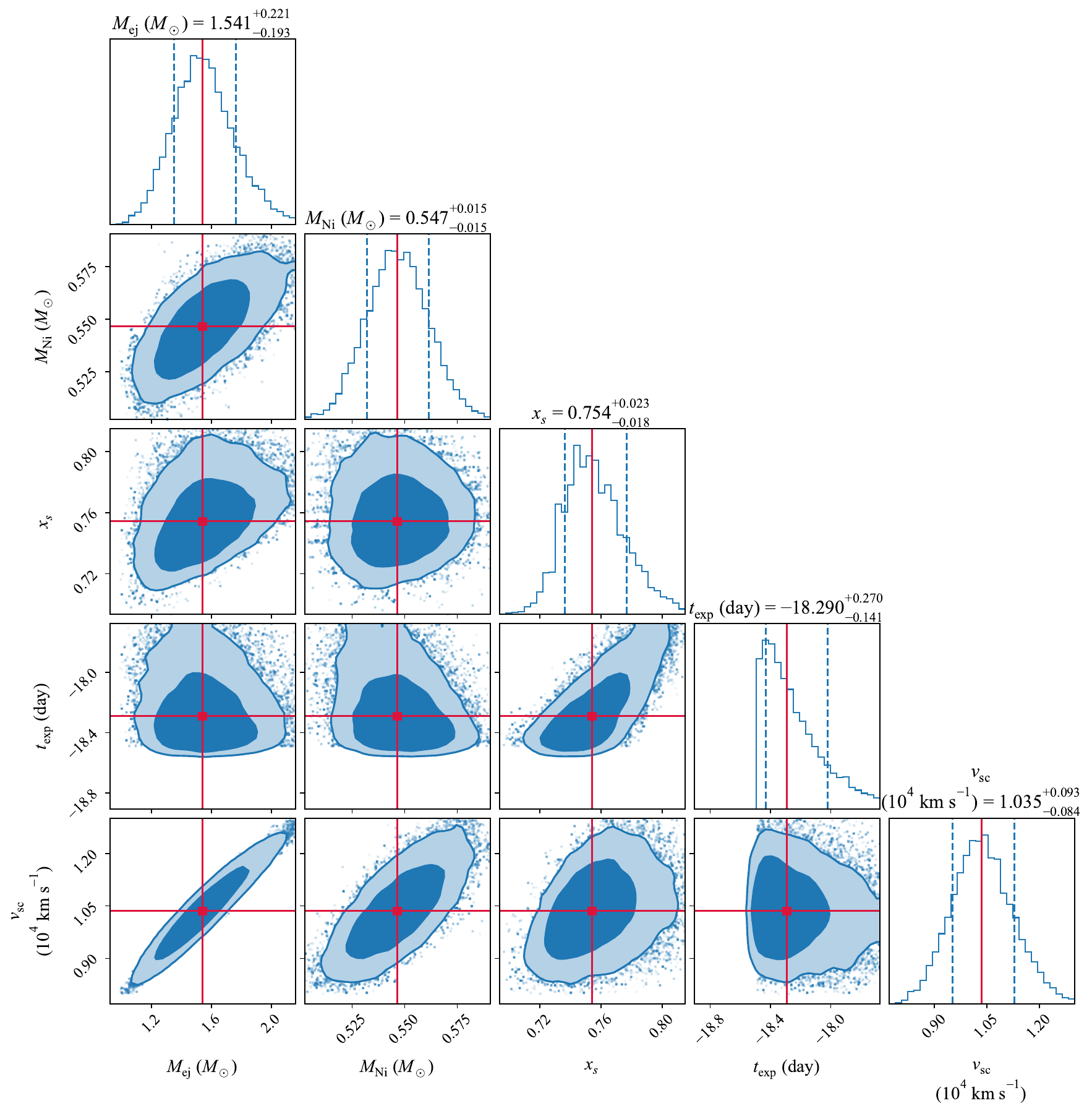}
        \caption{Corner plot showing the posterior probability distributions for the physical parameters of SN~2024igg obtained with our MCMC analysis using \textsc{Transfit}. The diagonal panels display the marginalized one-dimensional distributions for each, with vertical dashed lines and red solid lines indicating the central 68\% credible interval (1$\sigma$) and the median. Note that the uncertainty of $M_{\rm Ni}$ shown in the plot reflects only the fitting errors. When the distance uncertainty is incorporated, the final result yields $M_{\rm Ni} = 0.547^{+0.082}_{-0.082}$ $M_{\rm \odot}$.}
        \label{fig:corner}%
    \end{figure*}

The \textsc{Transfit} code \citep{2025ApJ...992...20L} was used to estimate the mass of $^{56}$Ni synthesized in the explosion and the total ejecta mass, assuming that the pseudobolometric luminosity of SN~2024igg is only due to the radioactive decay of $^{56}$Ni. \textsc{Transfit} is a novel framework that numerically solves a generalized energy-conservation equation, explicitly incorporating time-dependent radiative diffusion, continuous radioactive or
central-engine heating, and ejecta expansion dynamics. The $^{56}$Ni-powered model includes six parameters: ejecta mass $M_{\rm ej}$, kinetic energy $E_{\rm K}$ or scaled velocity $v_{\rm sc}$ ($E_{\rm K}=0.5M_{\rm ej}v_{\rm sc}^2$), $^{56}$Ni mass $M_{\rm Ni}$, maximum $^{56}$Ni distribution radius $x_{heat}$ (dimensionless within 0--1, in mass space), gray opacity $\kappa$, and opacity for gamma-ray
photons $\kappa_{\gamma}$. Constant opacities $\kappa=0.1$ cm$^2$ g $^{-1}$ \citep{2023MNRAS.521.1162D} and $\kappa_{\gamma}=0.025$ cm$^2$ \citep{2024MNRAS.533..994G} were adopted in the fitting. The explosion time $t_{\rm exp}$ relative to $B$-band maximum was included as an additional parameter, which is limited by the last nondetection and the first detection in ZTF $r$. The MCMC corner plot is shown in Fig.\,\ref{fig:corner}, which corresponds to $M_{\rm ej}=1.54^{+0.22}_{-0.19}$ $M_{\rm \odot}$, $M_{\rm Ni}=0.547^{+0.082}_{-0.082}$ $M_{\rm \odot}$ (distance error incorporated), $x_{\rm heat}=0.754^{+0.023}_{-0.018}$, $t_{\rm exp}=-18.29^{+0.27}_{-0.14}$ days, and $v_{\rm sc}=1.035^{+0.093}_{-0.084}\times10^4$ km s$^{-1}$.

\begin{table*}
\centering
\caption{Estimated pseudobolometric light curve of SN~2024igg.}\label{table:bolo_lc}
\begin{tabular}{lllllllll}
\hline
\hline
Phase\tablefootmark{a} & $L$ & Error\tablefootmark{b} & Phase & $L$ & Error & Phase & $L$ & Error \\
(Days)   & (10$^{42}$ erg s$^{-1}$) & (10$^{42}$ erg s$^{-1}$) & (days)   & (10$^{42}$ erg s$^{-1}$) & (10$^{42}$ erg s$^{-1}$) & (days)   & (10$^{42}$ erg s$^{-1}$) & (10$^{42}$ erg s$^{-1}$) \\
\hline
-13.31 & 2.48 & 0.13 & 1.82 & 11.91 & 0.61 & 36.17 & 2.56 & 0.11 \\
-11.48 & 4.90 & 0.27 & 6.16 & 10.12 & 0.60 & 37.11 & 2.41 & 0.13 \\
-9.08 & 8.20 & 0.48 & 12.28 & 7.13 & 0.25 & 46.05 & 1.85 & 0.06 \\
-7.06 & 10.51 & 0.58 & 20.19 & 4.69 & 0.38 & 48.94 & 1.68 & 0.05 \\
-5.05 & 12.22 & 0.66 & 25.17 & 3.82 & 0.09 & 58.79 & 1.35 & 0.06 \\
-4.14 & 12.85 & 0.56 & 30.30 & 3.21 & 0.09 & 70.72 & 1.01 & 0.03 \\
0.22 & 12.46 & 0.57 & 34.82 & 2.73 & 0.09 & - & - & - \\
\hline
\end{tabular}
\tablefoot{\\
  \tablefoottext{a}{Rest frame, with respect to the $B$-band maximum light, MJD$_{B\rm max}=60452.10$.}\\
  \tablefoottext{b}{Uncertainty in the distance not included. $1\sigma$.}}\\
\end{table*}

\end{appendix}

\end{document}